\documentclass[twocolumn,aps,prl,preprintnumbers]{revtex4}
\usepackage{bbm}
\usepackage[latin9]{inputenc}
\usepackage{amsmath}
\usepackage{amssymb}
\usepackage{graphicx}
\usepackage{mathrsfs}
\usepackage{amsfonts}
\usepackage{amsthm}
\usepackage{color}
\usepackage{txfonts}
\usepackage[colorlinks=true,citecolor=blue,linkcolor=blue,urlcolor=blue,anchorcolor=blue]{hyperref}%
\usepackage{pifont}
\hypersetup{colorlinks=true,citecolor=blue,linkcolor=blue,urlcolor=blue}
\providecommand{\U}[1]{\protect\rule{.1in}{.1in}}
\makeatletter
\@ifundefined{textcolor}{}
{
\definecolor{BLACK}{gray}{0}
\definecolor{WHITE}{gray}{1}
\definecolor{RED}{rgb}{1,0,0}
\definecolor{GREEN}{rgb}{0,1,0}
\definecolor{BLUE}{rgb}{0,0,1}
\definecolor{CYAN}{cmyk}{1,0,0,0}
\definecolor{MAGENTA}{cmyk}{0,1,0,0}
\definecolor{YELLOW}{cmyk}{0,0,1,0}
}

\makeatother
\begin{document}
\title{Realization of Wilson fermions in topolectrical circuits}
\author{Huanhuan Yang}
\author{Lingling Song}
\author{Yunshan Cao}
\author{Peng Yan}
\email[]{yan@uestc.edu.cn}
\affiliation{School of Electronic Science and Engineering and State Key Laboratory of Electronic Thin Films and Integrated Devices, University of
Electronic Science and Technology of China, Chengdu 610054, China}

\begin{abstract}
Wilson fermion (WF) is a fundamental particle in the theory of quantum chromodynamics, originally proposed by Kenneth Wilson to solve the fermion doubling problem, i.e., more fermions than expected when one puts fermionic fields on a lattice. In this Letter, we report a direct observation of the WF in circuit systems. It is found that WFs manifest as topological spin textures analogous to the half skyrmion, half-skyrmion pair, and N\'{e}el skyrmion structures, depending on their mass. Transformations of different WF states are realized by merely tuning the electric elements. Theoretical calculations have shown that the WF with a half-skyrmion profile represents a novel quantum anomalous semimetal phase supporting a chiral edge current [B. Fu \emph{et al.} \href{https://doi.org/10.1038/s41535-022-00503-0}{npj Quantum Mater. {\bf 7}, 94
(2022)}], but the experimental evidence is still lacking. We experimentally observe the propagation of chiral edge current along the domain-wall separating two circuits with contrast fractional Chern numbers. Our work presents the first experimental evidence for WFs in topolectrical circuits. The nontrivial analogy between the WF state and the skyrmionic structure builds an intimate connection between the two burgeoning fields.
\end{abstract}

\maketitle
Lattice quantum chromodynamics (QCD) is an effective method to study the strong interactions of quarks mediated by gluons \cite{Kogut1983,Gattringer2010}. In lattice QCD calculation, quarks are represented by fermionic fields and placed at lattice sites, and gluons play the role of interactions between neighboring sites \cite{DeTar2004}. However, when naively putting the fermionic fields on a lattice, we will meet the fermion doubling problem \cite{Nielsen1981}, i.e., the emergence of $2^d-1$ spurious fermionic particles for each original fermion ($d$ is the dimension of the spacetime). The origin of the doubling problem is deeply connected with chiral symmetry and can be traced back to the axial anomaly \cite{Chandrasekharan2004}. To remove the ambiguity, Kenneth Wilson developed a technique by introducing wave-vector-dependent mass, which modifies the Dirac fermions to Wilson ones \cite{Wilson1974}. The fermion doubling issue exists in condensed matter physics as well \cite{Semenoff1984,Messias2017,YangZ2021,Haldane1988,Yu2019}. It prevents the occurrence of quantum anomalies in lattices, such as the quantum anomalous Hall insulator \cite{Haldane1988} and Weyl semimetal with single node \cite{Yu2019}. It is known that Dirac fermions manifest as the low-energy excitations of topological semimetals/insulators (e.g. graphene) \cite{Neto2009,Qi2011,Armitage2018,Lv2021}. However, the observation of Wilson fermion (WF) is yet to be realized.

In this Letter, we utilize a lattice model to realize the WF and probe it in topolectrical circuit experiments  \cite{Jia2015,Albert2015,ZhaoAP2018,Imhof2018,Lee2018,Hadad2018,Hofmann2019,YWang2020,Helbig2020,RChen2020,Olekhno2020,Song2020,Yang2020,
Zhang2020,Yang2021,Yang2022,Ventra2022,Yang2022prb,Song2022,NEWu2022}. Interestingly enough, we find that the nontrivial state of the WF strongly depends on its mass and can be classified into three categories characterized by different Chern numbers of 0, $\pm$1/2, and $\pm$1, corresponding to the half-skyrmion pair, half skyrmion, and N\'{e}el skyrmion, respectively. We propose a circuit method to efficiently manipulate the transport and transformation of the WF states. In this system, the fractional Chern number dictates a novel quantum anomalous semimetal (QASM) phase with a chiral edge current as suggested in Ref. \cite{Fu2022}. Here, we report a direct observation of the chiral current along the domain wall (DW) separating two circuits with contrast fractional Chern numbers being $1/2$ and $-1/2$. WFs in a three-dimensional (3D) circuit system are constructed as well. They are characterized by 3D winding numbers and accompanied by the emergence of the surface states and DW states at the boundaries. Our work opens the door for realizing the exotic WFs in solid-state systems.

We begin from the Dirac Hamiltonian $\mathcal{H}=c{\bf k}\cdot\alpha+mc^2\beta$ with $c$ the light speed, {\bf k} the wave vector, and $\alpha,\beta$ being the Dirac matrices, which describes a Dirac fermion with the mass $m$ \cite{Dirac1928}. Expressing this Dirac Hamiltonian on a lattice of the tight-binding form, we obtain
$\mathcal{H}_{\rm D}=\sum^d_{i=1}\frac{\hbar v}{a}\sin(k_ia)\alpha_i+m v^2 \beta$ with $a$ the lattice constant and $\hbar v$ the hopping strength ($d$ is the space dimension). It is straightforward to verify that $2^d-1$ non-physics fermion doublers appear at the Brillouin zone (BZ) boundaries $k_i=\pi/a$. Following Wilson's method, we derive the WF Hamiltonian of the following form $\mathcal{H}=\mathcal{H}_{\rm D}+\mathcal{H}_{\rm W}$ with $\mathcal{H}_{\rm W}=\frac{4b}{a^2}\sin^2\frac{k_ia}{2}\beta$ being the {\bf k}-dependent Wilson mass term \cite{Fu2022}. Here, the {\bf k}-independent mass $m$ in Hamiltonian $\mathcal{H}_{\rm D}$ is referred to as the dispersionless mass of WFs. It is noted that the $\mathcal{H}_{\rm W}$ term breaks the parity symmetry in two-dimensional (2D) and chiral symmetry in 3D cases, which can circumvent the fermion doubling problems \cite{Nielsen1981} and reproduce the quantum anomaly in the continuum limit. In Supplemental Material \cite{SM}, we show the details how the doublers from Dirac Hamiltonian are removed by introducing the Wilson mass term. Next, we report the realization of the Wilson Hamiltonian in electrical circuits.

\begin{figure}
  \centering
  \includegraphics[width=0.48\textwidth]{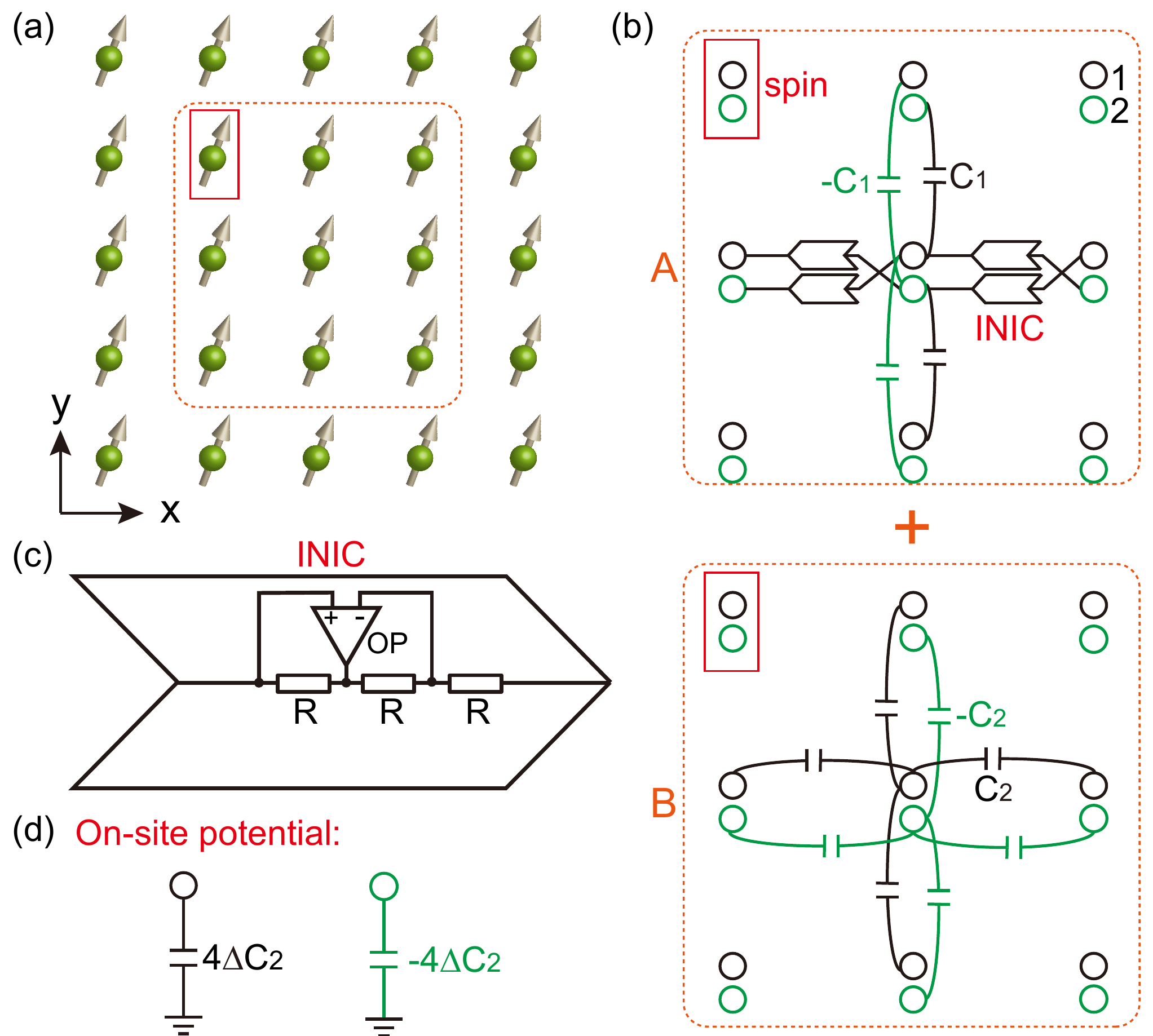}\\
  \caption{(a) Illustration of a 2D spinful square lattice. (b) The circuit realization of the hopping terms by A+B parts. Part A consists of two kinds of capacitors $\pm C_1$ and the INICs. Part B is composed of two types of capacitors $\pm C_2$. The red rectangle indicates the correspondence between spin and circuit nodes. (c) The details of INIC. The INIC is composed of an operational amplifier (OP) and three resistors $R$, acting as a positive (negative) resistor from right to left (left to right). (d) The realization of the staggered on-site potentials.}\label{2DM}
\end{figure}

We consider a 2D spinful square lattice in Fig. \ref{2DM}(a). The circuit is constituted by four types of capacitors $\pm C_{1,2}$ and the negative impedance converters with current inversion (INICs) in Fig. \ref{2DM}(b), where A and B parts correspond to the massless Dirac and Wilson mass Hamiltonians, respectively. It is noted that one can utilize inductors to replace negative capacitors because the admittance of the negative capacitor $-i\omega C$ is equivalent to the inductor $-i\frac{1}{\omega L}$ for $L=1/(C\omega^2)$ \cite{Zheng2022}. Here, $\omega$ is the working frequency. We implement two sites to imitate a (pseudo-) spin, indicated by the red rectangle in Figs. \ref{2DM}(a) and \ref{2DM}(b). The INIC is set up by an operational amplifier and three identical resistors $R$, as shown in Fig. \ref{2DM}(c). In Fig. \ref{2DM}(d), we show the realization of the staggered on-site potential, which models the dispersionless mass of WF.

The circuit response is governed by Kirchhoff's law $I(\omega)=\mathcal{J}(\omega)V(\omega)$ with $I$ the input current and $V$ the node voltage. The circuit Laplacian reads
\begin{equation}
\mathcal{J(\omega,{\bf k})}=\left(
                      \begin{array}{cc}
                        j_{11} & j_{12} \\
                        j_{21} & j_{22} \\
                      \end{array}
                    \right),
\end{equation}
where
$j_{11}=4i\omega C_2-2i\omega C_2(\cos k_x+\cos k_y)+4i \omega \Delta C_2,
j_{12}=2iG\sin k_x+2\omega C_1\sin k_y,
j_{21}=2iG\sin k_x-2\omega C_1\sin k_y,$ and
$j_{22}=-4i\omega C_2+2i\omega C_2(\cos k_x+\cos k_y)
-4i \omega \Delta C_2$, with $G=1/R$ the conductance and $\Delta$ being the mass coefficient of WFs. In the presence of conductance, the time-reversal symmetry ($\mathcal{T}$) of the system is broken because of $\mathcal{J}(\omega,{\bf k})^*\neq-\mathcal{J}(\omega,-{\bf k})$ \cite{Hofmann2019}.

By expressing $\mathcal{J(\omega)}=i\mathcal{H(\omega)}$ with the Dirac matrices, we obtain
\begin{equation}\label{H}
\begin{aligned}
\mathcal{H(\omega)}=&2G\sin k_x\alpha_x+2 \omega C_1\sin k_y\alpha_y\\
&+4\omega C_2(\sin^2{\frac{k_x}{2}}+\sin^2{\frac{k_y}{2}})\beta+4 \omega \Delta C_2\beta,
\end{aligned}
\end{equation}
where $\alpha_x$, $\alpha_y$ and $\beta$ represent the Pauli matrices $\sigma_x$, $\sigma_y$, and $\sigma_z$, respectively. It is noted that the above Hamiltonian fully simulates the lattice model of WFs. The first three terms in Eq. \eqref{H} represent the Hamiltonian of WF, and the last term is the dispersionless mass of WF. Meanwhile, by tuning the electric elements parameters $G$, $C_1$, and $C_2$, one can conveniently manipulate the shape of Wilson cones (similar to the Dirac cones).

In what follows, we analyze the topological properties of the lattice model.

For a $\mathcal{T}$-broken 2D two-band system, one can evaluate the Chern number \cite{Asbothbook}
\begin{equation}
\mathcal{C}=-\frac{1}{2\pi}\int_{\rm BZ}\left(\frac{\partial A_y}{\partial k_x}-\frac{\partial A_x}{\partial k_y}\right)dk_xdk_y,
\end{equation}
to judge its topological properties. Here ${\bf A}({\bf k})=i\left<u_{\bf k}\vert \nabla_{\bf k} \vert u_{\bf k}\right>$ is the Berry connection with $\left.\vert u_{\bf k}\right>$ the eigenstate of lower band.

\begin{figure*}[t!]
  \centering
  \includegraphics[width=0.98\textwidth]{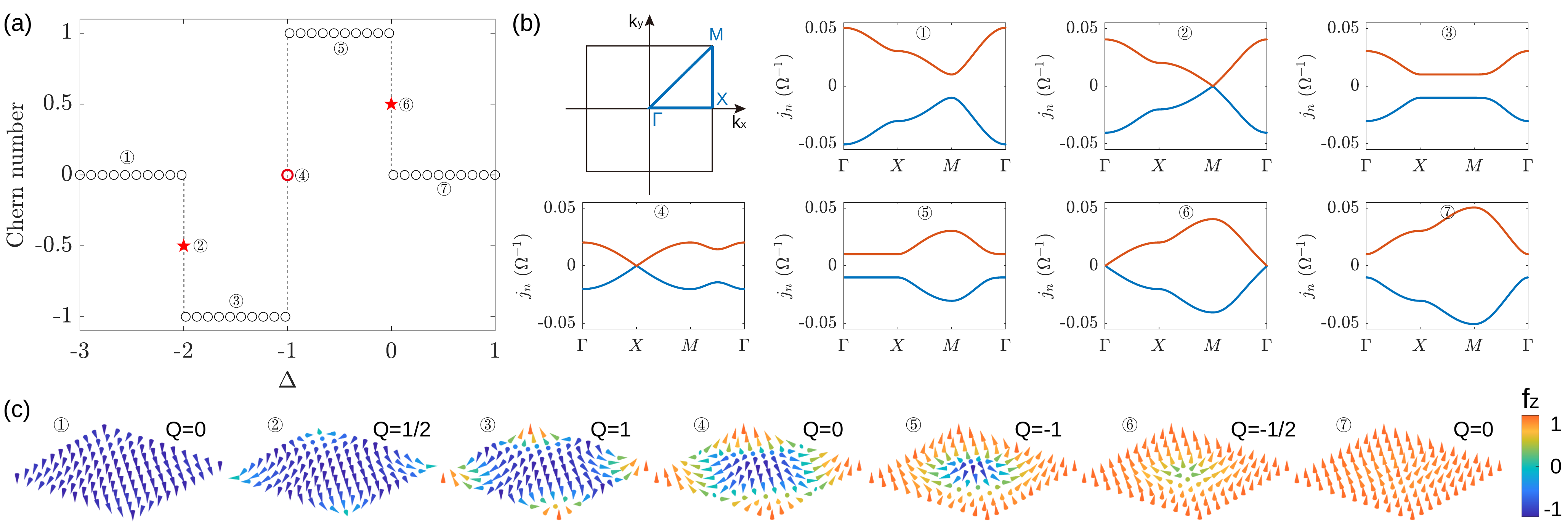}\\
  \caption{(a) The Chern number steps as a function of the dispersionless mass $\Delta$. (b) The first BZ and the typical band structures for $\Delta=-2.5,-2,-1.5,-1,-0.5,0,0.5$, respectively, corresponding to different parameter intervals in (a). (c) The spin textures of WFs in the momentum space.}\label{Tindex}
\end{figure*}

In following calculations, we adopt $C_i=C=1$ nF ($i=1,2$), $f=\omega/(2\pi)=806$ kHz (In experiments, we will use $L=39$ $\mu$H to replace $-C$, so we choose $\omega=1/\sqrt{LC}$), and $G=\omega C=0.005$ $\Omega^{-1}$ ($R=200$ $\Omega$). Calculations of Chern number as a function of the dispersionless mass parameter $\Delta$ are plotted in Fig. \ref{Tindex}(a). We find the Chern numbers are quantized to five values $\pm1$, $\pm \frac{1}{2}$, and 0. It is noted that the Chern number is irrelevant to the value of $C_1$ but becomes opposite if $C_2$ changes its sign. We show the first BZ and typical band structures in Fig. \ref{Tindex}(b). Combined with the topological index, we classify these topological phases as follows. The band gaps open for the parameter intervals \ding{172} and \ding{178}, where the Chern number is zero. It gives the trivial insulator phase. For parameters in \ding{173} and \ding{177}, the band gaps close at $M$ and $\Gamma$ points, respectively. Surprisingly, the Chern numbers are quantized to $\mp1/2$ , respectively. This novel phase is dubbed as the QASM \cite{Fu2022}. For parameter zones \ding{174} and \ding{176}, the band gaps open with the topological number $\mathcal{C}=\mp1$, both of which represent Chern insulators with opposite chiralities. For the parameter region \ding{175}, the band structure closes at $X$ point with a vanishing Chern number, indicating a normal semimetal phase.

By expressing Eq. \eqref{H} as $\mathcal{H}={\bf f}({\bf k})\cdot{\boldsymbol\sigma}$, one can define a unit spin vector $\hat{\bf f}({\bf k})$ as $\hat{{\bf f}}({\bf k})=\frac{\bf f ({\bf k})}{\vert{\bf f ({\bf k})}\vert}$, where ${\bf f}({\bf k})=(f_x,f_y,f_z)$ is the coefficient of Pauli matrices and $\vert{\bf f ({\bf k})}\vert=\sqrt{f_x^2+f_y^2+f_z^2}$. Figure \ref{Tindex}(c) displays the spin textures of 2D WFs, which evolve as the increasing of dispersionless mass. The spin textures are reminiscent of the magnetic solitons in the condensed matter system  \cite{Tokura2021,Yu2021,Chen2021}. It is observed that the spin textures of trivial insulator, Chern insulator, QASM, and normal semimetal correspond to the ferromagnetic ground state, skyrmion, half skyrmion, and half-skyrmion pair, respectively. By evaluating the topological charge $Q=\frac{1}{4\pi}\int_{\rm BZ} \hat{\bf f}\cdot(\frac{\partial \hat{\bf f}}{\partial k_x}\times \frac{\partial \hat{\bf f}}{\partial k_y}) dk_xdk_y$ in Fig. \ref{Tindex}(c), we identify an intimate connection with the Chern number as $Q+\mathcal{C}=0$. This finding thus establishes an interesting map between WFs and magnetic solitons in electrical circuits.

In the broad spintronics community, the manipulation of skyrmion motion is crucial for the next-generation information industry \cite{Fert2017}. Here, we propose a method to control the circuit skyrmion motion in momentum space. We first consider a skyrmion configuration with $\Delta=-0.5$. To generate a skyrmion propagation along $k_x$ direction over a distance $k_0$, one can modify Eq. \eqref{H} to
$\mathcal{H(\omega)}=2G\sin (k_x-k_0)\sigma_x+2 \omega C_1\sin k_y\sigma_y-4\omega C_2[\cos(k_x-k_0)+\cos k_y+\frac{1}{2}]\sigma_z$, which can be recast as $\mathcal{H(\omega)}=2G'\sin k_x\sigma_x-2G''\cos k_x\sigma_x+2 \omega C_1\sin k_y\sigma_y-[4\omega C'_2\cos k_x+4\omega C''_2\sin k_x+4\omega C_2(\cos k_y+\frac{1}{2})]\sigma_z$. Compared with the original Eq. \eqref{H}, one merely needs to modify two hopping strengths ($G'=G\cos k_0$ and $C_2'=C_2\cos k_0$) and to add two extra hopping terms ($-2G''\cos k_x\sigma_x$ and $4\omega C''_2\sin k_x\sigma_z$). Variable resistors and capacitors can be conveniently adopted to realize these operations in circuit devices.

\begin{figure*}
  \centering
  \includegraphics[width=0.98\textwidth]{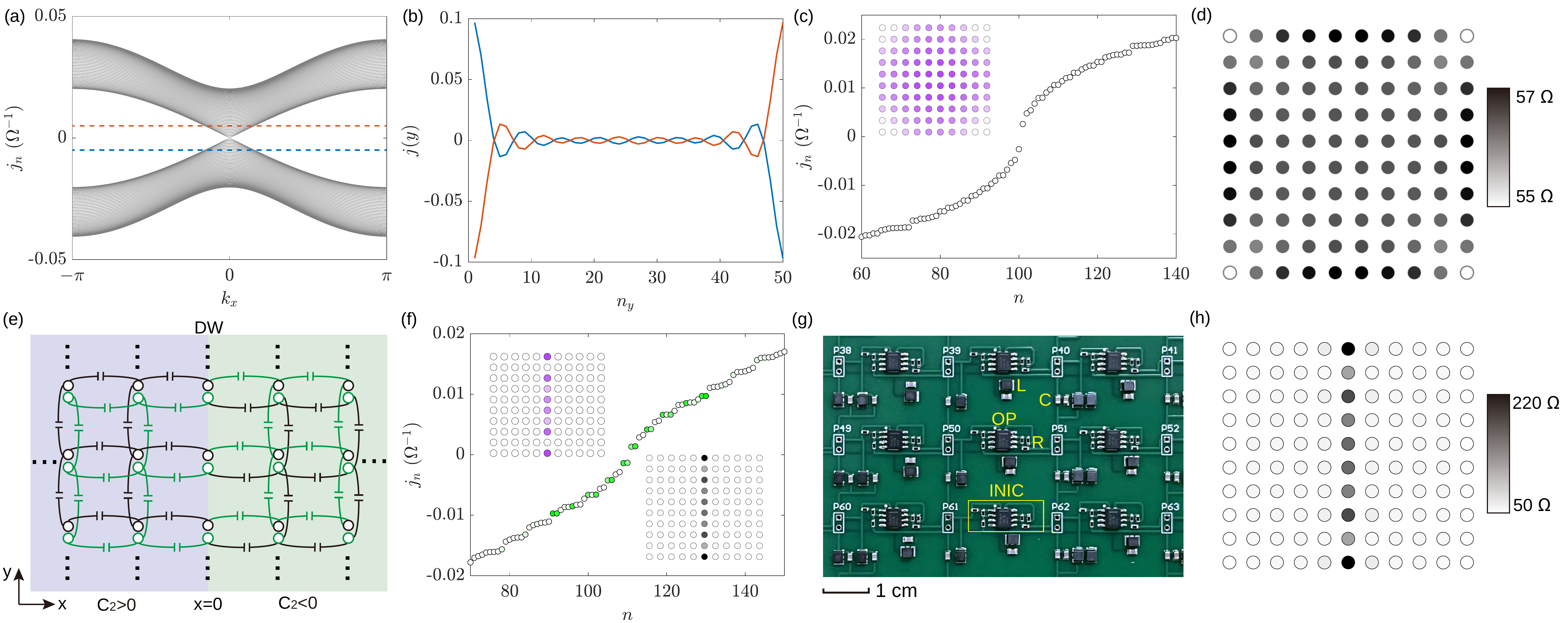}\\
  \caption{(a) Admittance spectrum of the ribbon geometry. (b) The current density distribution for two different ``Fermi" levels slightly deviating from the $j_n=0$ $\Omega^{-1}$ [dashed lines in (a)]. (c) The admittance spectrum with the inset showing the wave functions near $j_n=0$ $\Omega^{-1}$. (d) The impedance distribution of the sample. (e) The configuration of the circuit DW with $\pm 1/2$ topological charges in light blue and green regions. (f) The admittance spectrum. Insets: The distribution of the wave functions and impedances. (g) The partial printed circuit board used in the experiment. (h) Experimentally measured impedance.}\label{QAS}
\end{figure*}

To show the properties of QASM ($\Delta=0$ with $\mathcal{C}=\frac{1}{2}$), we consider a ribbon configuration with periodic boundary condition along $\hat{x}$ direction and $N_y=50$ nodes along $\hat{y}$ direction. Figure \ref{QAS}(a) shows the admittance spectra, where the conduction and valence bands touch at $k_x=0$. There is no isolated band in this admittance spectra, so the edge state is absent. Considering the Hamiltonian \eqref{H}, one can define a velocity operator as $v=1/(i\hbar)[x,\mathcal{H}]=\frac{\partial \mathcal{H}}{\partial k_x}=2G\cos k_x\sigma_x+2\omega C_2\sin k_x\sigma_z$. The transverse current density then can be written as
\begin{equation} \label{jd}
\begin{aligned}
j(y)=\sum_n^{\varepsilon_n<\mu}\int&\Big [2G\cos k_x\phi^\dagger_n(k_x,y)\sigma_x\phi_n(k_x,y)\\
&+2\omega C_2\sin k_x\phi^\dagger_n(k_x,y)\sigma_z\phi_n(k_x,y)\Big ]dk_x,
\end{aligned}
\end{equation}
with $\phi_n(k_x,y)$ being the wave functions of the $n$-th band and the sum index $n$ indicating the bands below the admittance $\mu$.
We plot the current density for different positions in Fig. \ref{QAS}(b). It is found that the current density decays from the boundary nodes, and its values are opposite for the top and bottom edges. Consequently, the chiral edge currents constitute the new bulk-edge correspondence of QASM \cite{Fu2022,Zou2022}.

Then, we consider a $10\times10$ square lattice to study the finite-size effect. Diagonalizing the corresponding circuit Laplacian, we obtain the admittance spectra shown in Fig. \ref{QAS}(c) and the wave functions near $j_n=0$ $\Omega^{-1}$ in the inset of Fig. \ref{QAS}(c). In our circuit, the impedance between the node $a$ and the ground is computed by $Z_{a,{\rm ground}}=\sum_{n}\frac{\vert\phi_{n,a}\vert^2}{j_n}$ with $\phi_{n,a}$ the wave function of node $a$ for $n$th admittance mode, which reflects the features of wave functions near $j_n=0$ $\Omega^{-1}$ and can be measured readily  \cite{Lee2018,Yang2020}. By comparing Figs. \ref{QAS}(d) and \ref{QAS}(c) (inset), we find that the impedance of each node against the ground exhibits almost the same spatial distribution to wave functions, and one does not find an edge state as expected.

To demonstrate the bulk-boundary correspondence, we consider a one-dimensional DW with $10\times11$ ``spins" (an extra column is set up for DW configuration), as shown in Fig. \ref{QAS}(e), where the capacitor $C_2$ has a kink at the center of the sample, i.e., $C_2>0$ ($<0$) in the light blue (green) region. In this circumstance, the Chern number varies from $1/2$ (left) to $-1/2$ (right). The eigenvalue and wavefunction of the bound state can be solved as $J=2\omega C_1k_y$ and $\phi(x,y)=\frac{1}{\sqrt{2\pi}}\chi_y\sqrt{\lambda}
\exp(-\lambda \vert x\vert+ik_yy)$ with $\chi_y=\frac{\sqrt{2}}{2}(-i,1)^{\rm T}$ and $\lambda=\frac{C_1}{\vert C_2 \vert}+\sqrt{(\frac{C_1}{C_2})^2+k_y^2}$. One can obtain the effective velocity of the bound state $v_{\rm eff}=\frac{\partial J}{\partial k_y}=2\omega C_1$, indicting the DW bound state propagating along $\hat{y}$ direction \cite{SM}. In Fig. \ref{QAS}(f), we show the admittance spectrum with the insets displaying the wave functions and impedances, from which one can clearly see the bound state confined inside the DW. Here, we use the inverse participation ratio $p=\lg(\sum_i\vert\phi_{n,i}\vert^4)$ of the system to characterize the localization properties of the wave functions \cite{HAraki2019,Wakao2020}, and the green dots indicate the localized state.

Then, we prepare a printed circuit board to verify these theoretical predictions, as shown in Fig. \ref{QAS}(g) \cite{SM}. Figure \ref{QAS}(h) shows the experimental impedance distribution. It demonstrates a localized state between two domains, which compares well with the theoretical result in Fig. \ref{QAS}(f). The existence of one bound state is closely related to the fact that the topological invariants between the two sides of the DW differ by 1. In addition, one cannot observe the edge states on the rest boundaries of the sample, which confirms that there is indeed no edge state.

To show the chiral propagation of the bound state, we perform the circuit simulation with the software LT{\tiny SPICE} \cite{LT}. By inputting a Gauss signal close to the DW, we observe the bound mode propagating along the $\hat{y}$ direction of DW. Finally, the voltage signal becomes a steady bound-state inside the DW \cite{SM}.

To characterize Chern insulators ($\mathcal{C}=\pm1$), we compute the admittance for a ribbon configuration \cite{SM}. For the parameter $\Delta=-0.5$ and $-1.5$, we find two crossing bands in the admittance gaps but with opposite charities. To show the chiral propagation of edge states, we perform the circuit simulation on a finite-size square lattice and observe a chiral voltage propagation \cite{SM}.

\begin{figure}[b!]
  \centering
  \includegraphics[width=0.5\textwidth]{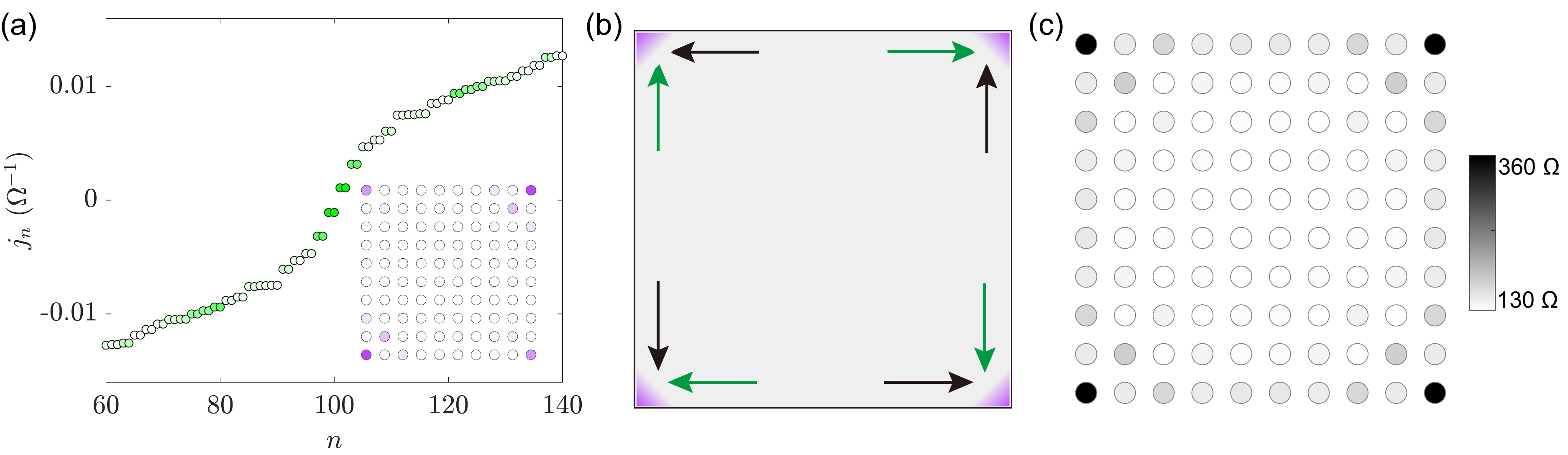}\\
  \caption{(a) The admittance of the finite-size square lattice with $10\times10$ ``spins" with the inset showing the wave functions near $j_n=0$ $\Omega^{-1}$. (b) The corner state formed by the convergence of Chern insulators with opposite chiralities. (c) Numerical impedance.}\label{Q0SM}
\end{figure}

Interestingly, we note that at the phase transition point separating two Chern insulators ($\Delta=-1$), the Chern number vanishes but the spin texture is still non-trivial. We consider a finite-size lattice with $10\times10$ ``spins". The admittance spectrum is plotted in Fig. \ref{Q0SM}(a), showing that a series of localized states lie near $j_n=0$ $\Omega^{-1}$. The wave functions of localized states are displayed in the inset of Fig. \ref{Q0SM}(a), from which we identify a corner state. The origin of the emerging corner states can be interpreted as the convergence of two Chern insulators with opposite chiralities, as shown by the green and black arrows in Fig. \ref{Q0SM}(b). Due to the contrast of the chirality, the one-dimensional edge states can only accumulate at the sample corners, forming the zero-dimensional localized states, i.e., corner states. These corner states can be detected by measuring the distributions of impedance, as shown in Fig. \ref{Q0SM}(c).

As a nontrivial generalization, we extend this model to a 3D system \cite{SM}. The circuit Hamiltonian then can be written as
\begin{equation}
\begin{aligned}
\mathcal{H}(\omega)=&2G\sin k_x \alpha_x+2\omega C_1\sin k_y\alpha_y
+2G\sin k_z\alpha_z\\
&+4 \omega C_2\left[\sin^2\frac{k_x}{2}+
\sin^2\frac{k_y}{2}+\sin^2\frac{k_z}{2}\right]\beta+4\Delta \omega C_2\beta,
\end{aligned}
\end{equation}
where $\alpha_x=\sigma_x\otimes\sigma_x$, $\alpha_y=\sigma_x\otimes\sigma_y$,
$\alpha_z=\sigma_x\otimes\sigma_z$, and $\beta=\sigma_z\otimes\sigma_0$ are Dirac matrices.

The topological properties of the 3D system are characterized by the winding number $w_3$ \cite{Schnyder2008}. Interestingly, we find that the topological index $w_3$ can only take five quantized values, i.e., 0, $\pm\frac{1}{2}$, $\pm 1$. For the topological insulator phase $w_3=1$, one can observe surface states. At the border of the two topological insulators with opposite winding numbers, we find the hinge states induced by the overlap of the surface states. For the QASM phase $w_3=1/2$, we observe the bounded surface state in a finite-size DW circuit along the $\hat{x}$ direction \cite{SM}.

To summarize, we experimentally observed WFs in circuit systems. In addition, we mapped WFs with different masses or configurations to magnetic solitons with different skyrmion charges, which will enable us to study the properties of skyrmions, half skyrmions, and half-skyrmion pairs in electrical circuit platforms. We showed that the nontrivial spin-texture of WFs in momentum space is fully characterized by Chern numbers and winding number in 2D and 3D systems, respectively. The chiral edge current associated with the novel QASM state dictated by a fractional Chern number was directly detected. Our work presents the first circuit realization of WFs, which sets a paradigm for other platforms, such as cold atoms, photonic, and phononic metamaterials, to further explore these fascinating phenomena.

\begin{acknowledgments}
This work was supported by the National Key Research Development Program under Contract No. 2022YFA1402802 and the National Natural Science Foundation of China (Grants No. 12074057, No. 11604041, and No. 11704060)
\end{acknowledgments}


\begin{thebibliography}{99}
\bibitem {Kogut1983}J. B. Kogut, The lattice gauge theory approach to quantum chromodynamics, \href{https://doi.org/10.1103/RevModPhys.55.775}{Rev. Mod. Phys. \textbf{55}, 775 (1983)}.

\bibitem {Gattringer2010}C. Gattringer and C. B. Lang, Quantum Chromodynamics on the Lattice, Lecture Notes in physics 788 (Springer, 2010).

\bibitem {DeTar2004}C. DeTar and S. Gottlieb, Lattice Quantum Chromodynamics Comes of Age, \href{https://doi.org/10.1063/1.1688069}{Phys. Today \textbf{57}, 45 (2004)}.

\bibitem {Nielsen1981}H. B. Nielsen and M. Ninomiya, A no-go theorem for regularizing chiral fermions, \href{https://doi.org/10.1016/0370-2693(81)91026-1}{Phys. Lett. B \textbf{105}, 219 (1981)}.

\bibitem {Chandrasekharan2004}A. Chandrasekharan and U.-J. Wiese, An introduction to chiral symmetry on the lattice, \href{https://doi.org/10.1016/j.ppnp.2004.05.003}{Prog. Part. Nucl. Phys. \textbf{53}, 373 (2004)}.

\bibitem {Wilson1974}K. G. Wilson, Confinement of quarks, \href{https://doi.org/10.1103/PhysRevD.10.2445}{Phys. Rev. D \textbf{10}, 2445 (1974)}.

\bibitem{Semenoff1984}G. W. Semenoff, Condensed-Matter Simulation of a Three-Dimensional Anomaly, \href{https://doi.org/10.1103/PhysRevLett.53.2449}{Phys. Rev. Lett. \textbf{53}, 2449 (1984)}.

\bibitem {Messias2017}B. Messias de Resende, F. Crasto de Lima, R. H. Miwa, E. Vernek, and G. J. Ferreira, Confinement and fermion doubling problem in Dirac-like Hamiltonians, \href{https://doi.org/10.1103/PhysRevB.96.161113}{Phys. Rev. B \textbf{96}, 161113(R) (2017)}.

\bibitem{YangZ2021}Z. Yang, A. P. Schnyder, J. Hu, and C.-K. Chiu, Fermion Doubling Theorems in Two-Dimensional Non-Hermitian Systems for Fermi Points and Exceptional Points, \href{https://doi.org/10.1103/PhysRevLett.126.086401}{Phys. Rev. Lett. \textbf{126}, 086401 (2021)}.


\bibitem{Haldane1988}F. D. M. Haldane, Model for a Quantum Hall Effect without Landau Levels: Condensed-Matter Realization of the ``Parity Anomaly", \href{https://doi.org/10.1103/PhysRevLett.61.2015}{Phys. Rev. Lett. \textbf{61}, 2015 (1988)}.

\bibitem{Yu2019}Z. Yu, W. Wu, Y. X. Zhao, and S. A. Yang, Circumventing the no-go theorem: A single Weyl point without surface Fermi arcs, \href{https://doi.org/10.1103/PhysRevB.100.041118}{Phys. Rev. B \textbf{100}, 041118(R) (2019)}.

\bibitem {Neto2009}A. H. Castro Neto, F. Guinea, N. M. R. Peres, K. S. Novoselov, and A. K. Geim, The electronic properties of graphene, \href{https://doi.org/10.1103/RevModPhys.81.109}{Rev. Mod. Phys. \textbf{81}, 109 (2009)}.

\bibitem {Qi2011}X.-L. Qi and S.-C. Zhang, Topological insulators and superconductors, \href{https://doi.org/10.1103/RevModPhys.83.1057}{Rev. Mod. Phys. \textbf{83}, 1057 (2011)}.

\bibitem {Armitage2018}N. P. Armitage, E. J. Mele, and A. Vishwanath, Weyl and Dirac semimetals in three-dimensional solids, \href{https://doi.org/10.1103/RevModPhys.90.015001}{Rev. Mod. Phys. \textbf{90}, 015001 (2018)}.

\bibitem {Lv2021}B. Q. Lv, T. Qian, and H. Ding, Experimental perspective on three-dimensional topological semimetals, \href{https://doi.org/10.1103/RevModPhys.93.025002}{Rev. Mod. Phys. \textbf{93}, 025002 (2021)}.

\bibitem{Jia2015} N. Jia, O. Clai, S. Ariel, S. David, and S. Jonathan, Time- and Site-Resolved Dynamics in a Topological Circuit, \href{https://doi.org/10.1103/PhysRevX.5.021031}{Phys. Rev. X \textbf{5}, 021031 (2015)}.

\bibitem{Albert2015} V. V. Albert, L. I. Glazman, and L. Jiang, Topological Properties of Linear Circuit Lattices, \href{https://doi.org/10.1103/PhysRevLett.114.173902}{Phys. Rev. Lett. \textbf{114}, 173902 (2015)}.

\bibitem{ZhaoAP2018}E. Zhao, Topological circuits of inductors and capacitors,
\href{https://doi.org/10.1016/j.aop.2018.10.006}{Ann. Phys. \textbf{399}, 289 (2018)}.

\bibitem{Imhof2018}S. Imhof, C. Berger, F. Bayer, J. Brehm, L. W. Molenkamp, T. Kiessling, F. Schindler, C. H. Lee, M. Greiter, T. Neupert, and R. Thomale, Topolectrical-circuit realization of topological corner modes, \href{https://doi.org/10.1038/s41567-018-0246-1}{Nat. Phys. \textbf{14}, 925 (2018)}.

\bibitem{Lee2018}C. H. Lee, S. Imhof, C. Berger, F. Bayer, J. Brehm, L. W. Molenkamp, T. Kiessling, and R. Thomale, Topolectrical Circuits, \href{https://doi.org/10.1038/s42005-018-0035-2}{Comm. Phys. \textbf{1}, 39 (2018)}.

\bibitem{Hadad2018}Y. Hadad, J. C. Soric, A. B. Khanikaev, and A. Al\'{u}, Self-induced
topological protection in nonlinear circuit arrays, \href{https://doi.org/10.1038/s41928-018-0042-z}{Nat. Electron. \textbf{1}, 178 (2018)}.

\bibitem{Hofmann2019}T. Hofmann, T. Helbig, C. H. Lee, M. Greiter, and R. Thomale, Chiral Voltage Propagation and Calibration in a Topolectrical Chern Circuit, \href{https://doi.org/10.1103/PhysRevLett.122.247702}{Phys. Rev. Lett. \textbf{122}, 247702 (2019)}.

\bibitem{YWang2020}Y. Wang, H. M. Price, B. Zhang, and Y. D. Chong, Circuit
implementation of a four-dimensional topological insulator, \href{https://doi.org/10.1038/s41467-020-15940-3}{Nat. Commun. \textbf{11}, 2356 (2020)}.

\bibitem{Helbig2020}T. Helbig, T. Hofmann, S. Imhof, M. Abdelghany, T. Kiessling, L. W. Molenkamp, C. H. Lee, A. Szameit, M. Greiter, and R. Thomale, Generalized bulk-boundary correspondence in non-Hermitian topolectrical circuits, \href{https://doi.org/10.1038/s41567-020-0922-9}{Nat. Phys. \textbf{16}, 747 (2020)}.


\bibitem{RChen2020}R. Chen, C.-Z. Chen, J.-H. Gao, B. Zhou, and D.-H. Xu, Higher-Order Topological Insulators in Quasicrystals, \href{https://doi.org/10.1103/PhysRevLett.124.036803}{Phys. Rev. Lett. \textbf{124}, 036803 (2020)}.

\bibitem{Olekhno2020}N. A. Olekhno, E. I. Kretov, A. A. Stepanenko, P. A. Ivanova, V. V. Yaroshenko, E. M. Puhtina, D. S. Filonov, B. Cappello, L. Matekovits, and M. A. Gorlach, Topological edge states of interacting photon pairs emulated in a topolectrical circuit, \href{https://doi.org/10.1038/s41467-020-14994-7}{Nat. Commun. \textbf{11}, 1436 (2020)}.


\bibitem{Song2020}L. Song, H. Yang, Y. Cao, and P. Yan, Realization of the square-root higher-order topological insulator in electric circuits, \href{https://doi.org/10.1021/acs.nanolett.0c03049}{Nano Lett. \textbf{20}, 7566 (2020)}.

\bibitem{Yang2020}H. Yang, Z.-X. Li, Y. Liu, Y. Cao, and P. Yan, Observation of symmetry-protected zero modes in topolectrical circuits, \href{https://doi.org/10.1103/PhysRevResearch.2.022028}{Phys. Rev. Res. \textbf{2}, 022028(R) (2020)}.

\bibitem{Zhang2020}X.-X. Zhang and M. Franz, Non-Hermitian Exceptional Landau Quantization in Electric Circuits, \href{https://doi.org/10.1103/PhysRevLett.124.046401}{Phys. Rev. Lett. \textbf{124}, 046401 (2020)}.

\bibitem{Yang2021}H. Yang, L. Song, Y. Cao, X. R. Wang, and P. Yan, Experimental observation of edge-dependent quantum pseudospin Hall effect, \href{https://doi.org/10.1103/PhysRevB.104.235427}{Phys. Rev. B \textbf{104}, 235427 (2021)}.

\bibitem {Yang2022}H. Yang, L. Song, Y. Cao, and P. Yan, Experimental Realization of Two-Dimensional Weak Topological Insulators, \href{https://doi.org/10.1021/acs.nanolett.2c00555}{Nano. Lett. \textbf{22}, 3125 (2022)}.

\bibitem {Ventra2022}M. D. Ventra, Y. V. Pershin, and C.-C. Chien, Custodial Chiral Symmetry in a Su-Schrieffer-Heeger Electrical Circuit with Memory, \href{https://doi.org/10.1103/PhysRevLett.128.097701}{Phys. Rev. Lett. \textbf{128}, 097701 (2022)}.

\bibitem {Yang2022prb}H. Yang, L. Song, Y. Cao, and P. Yan, Observation of type-III corner states induced by long-range interactions, \href{https://doi.org/10.1103/PhysRevB.106.075427}{Phys. Rev. B \textbf{106}, 075427 (2022)}.

\bibitem{Song2022}L. Song, H. Yang, Y. Ca, and P. Yan, Square-root higher-order Weyl semimetals, \href{https://doi.org/10.1038/s41467-022-33306-9}{Nat. Commun. \textbf{13}, 5601 (2022)}.

\bibitem{NEWu2022}J. Wu, Z. Wang, Y. Biao, F. Fei, S. Zhang, Z. Yin, Y. Hu, Z. Song, T. Wu, F. Song, and R. Yu, Non-Abelian gauge fields in circuit systems, \href{https://doi.org/10.1038/s41928-022-00833-8}{Nat. Electron. \textbf{5}, 635 (2022)}.


\bibitem {Fu2022}B. Fu, J.-Y. Zou, Z.-A. Hu, H.-W. Wang, and S.-Q. Shen, Quantum Anomalous Semimetals, \href{https://doi.org/10.1038/s41535-022-00503-0}{npj Quantum Materials \textbf{7}, 94 (2022)}.

\bibitem {Dirac1928}P. A. M. Dirac, The Quantum Theory of the Electron, \href{https://doi.org/10.1098/rspa.1928.0023}{Proc. R. Soc. A \textbf{117}, 610 (1928)}.


\bibitem {SM}See Supplemental Material at http://link.aps.org/
supplemental/ for (I) the solution to the fermion doubling problem, (II) the propagation of DW bound state, (III) experimental details, (IV) Chern insulators, and (V) three-dimensional Wilson fermions.

\bibitem {Zheng2022}X. Zheng, T. Chen, and X. Zhang, Topolectrical circuit realization of quadrupolar surface semimetals, \href{https://doi.org/10.1103/PhysRevB.106.035308}{Phys. Rev. B \textbf{106}, 035308 (2022)}.

\bibitem {Asbothbook}J. K. Asb\'{o}th , L. Oroszl\'{a}ny, and A. P\'{a}lyi, A short course on topological insulator, Lecture Notes in physics 919 (Springer, 2016).

\bibitem {Yu2021}H. Yu, J. Xiao, and H. Schultheiss, Magnetic texture based magnonics, \href{https://doi.org/10.1016/j.physrep.2020.12.004}{Phys. Rep. \textbf{905}, 1 (2021)}.


\bibitem {Chen2021}S. Chen, S. Yuan, Z. Hou, Y. Tang, J. Zhang, T. Wang, K. Li,  W. Zhao, X. Liu, L. Chen, L. W. Martin, and Z. Chen, Recent Progress on Topological Structures in Ferroic Thin Films and Heterostructures, \href{https://doi.org/10.1002/adma.202000857}{Adv. Mater. \textbf{33}, 2000857 (2021)}.

\bibitem {Tokura2021}Y. Tokura and N. Kanazawa, Magnetic Skyrmion Materials,  \href{https://doi.org/10.1021/acs.chemrev.0c00297}{Chem. Rev. \textbf{121}, 2857 (2021)}.


\bibitem {Fert2017}A. Fert, N. Reyren, and V. Cros, Magnetic skyrmions: advances in physics and potential applications, \href{https://doi.org/10.1038/natrevmats.2017.31}{Nat. Rev. Mater. \textbf{2}, 17031 (2017)}.

\bibitem {Zou2022}J.-Y. Zou, B. Fu, H.-W. Wang, Z.-A. Hu, and S.-Q. Shen, Half-quantized Hall effect and power law decay of edge-current distribution, \href{https://doi.org/10.1103/PhysRevB.105.L201106}{Phys. Rev. B \textbf{105}, L201106 (2022)}.


\bibitem {HAraki2019}H. Araki, T. Mizoguchi, and Y. Hatsugai, Phase diagram of a disordered higher-order topological insulator: A machine learning study, \href{https://doi.org/10.1103/PhysRevB.99.085406}{Phy. Rev. B \textbf{99}, 085406 (2019)}.

\bibitem {Wakao2020}H. Wakao, T. Yoshida, H. Araki, T. Mizoguchi, and Y. Hatsugai, Higher-order topological phases in a spring-mass model on a breathing kagome lattice, \href{https://doi.org/10.1103/PhysRevB.101.094107}{Phys. Rev. B \textbf{101}, 094107 (2020)}.

\bibitem {LT} http://www.linear.com/LTspice.

\bibitem{Schnyder2008}A. P. Schnyder, S. Ryu, A. Furusaki, and A. W. W. Ludwig, Classification of topological insulators and superconductors in three spatial dimensions, \href{https://doi.org/10.1103/PhysRevB.78.195125}{Phys. Rev. B \textbf{78}, 195125 (2008)}.
\end{thebibliography}

\begin{thebibliography}{99}
\bibitem{Schnyder2008}A. P. Schnyder, S. Ryu, A. Furusaki, and A. W. W. Ludwig, Classification of topological insulators and superconductors in three spatial dimensions, \href{https://doi.org/10.1103/PhysRevB.78.195125}{Phys. Rev. B \textbf{78}, 195125 (2008)}.

\bibitem {Shenbook}S. Q. Shen, Topological insulators: Dirac equation in condesed matters, Springer Series of Solid State Science, Vol. 174 (Springer, 2012).

\bibitem {Shan2010} W.-Y. Shan, H.-Z. Lu, and S.-Q. Shen, Effective continuous model for surface states and thin films of three-dimensional topological insulators, \href{https://doi.org/10.1088/1367-2630/12/4/043048}{New J. Phys. \textbf{12}, 043048 (2010)}.
\end{thebibliography}

\newpage
\begin{widetext}

\begin{flushleft}
\center{\LARGE{\textbf{Supplemental Material:}}}
\\[0.5cm]

{\Large{\textbf{Realization of Wilson fermions in topolectrical circuits}}}
\quad\par
\quad\par
Huanhuan Yang, Lingling Song, Yunshan Cao,  and Peng Yan
\quad\par
\quad\par

\emph{\small{School of Electronic Science and Engineering and State Key Laboratory of Electronic Thin Films and Integrated Devices, University of Electronic Science and Technology of China, Chengdu 610054, China}}

\end{flushleft}

\section{I. The solution to the fermion doubling problem}
In this section, we show how to remove the doublers of Dirac fermions. In the main text, we have expressed the massless Dirac and Wilson Hamiltonians as
$\mathcal{H}_{\rm D}=\sum^d_{i=1}\frac{\hbar v}{a}\sin(k_ia)\alpha_i$ and $\mathcal{H}=\mathcal{H}_{\rm D}+\mathcal{H}_{\rm W}$ with $\mathcal{H}_{\rm W}=\frac{4b}{a^2}\sin^2\frac{k_ia}{2}\beta$, respectively. In Figs. \ref{BS}(a) and \ref{BS}(b), we display the band structures of the 2D square and 3D hyper-cubic lattices. The green bands indicate the Dirac fermions with 3 and 7 doublers (gray dots in the first Brillouin zones), and the blue and red bands represent the Wilson fermions only appearing at the $\Gamma$ point.
\begin{figure}[h!]
  \centering
  \includegraphics[width=1\textwidth]{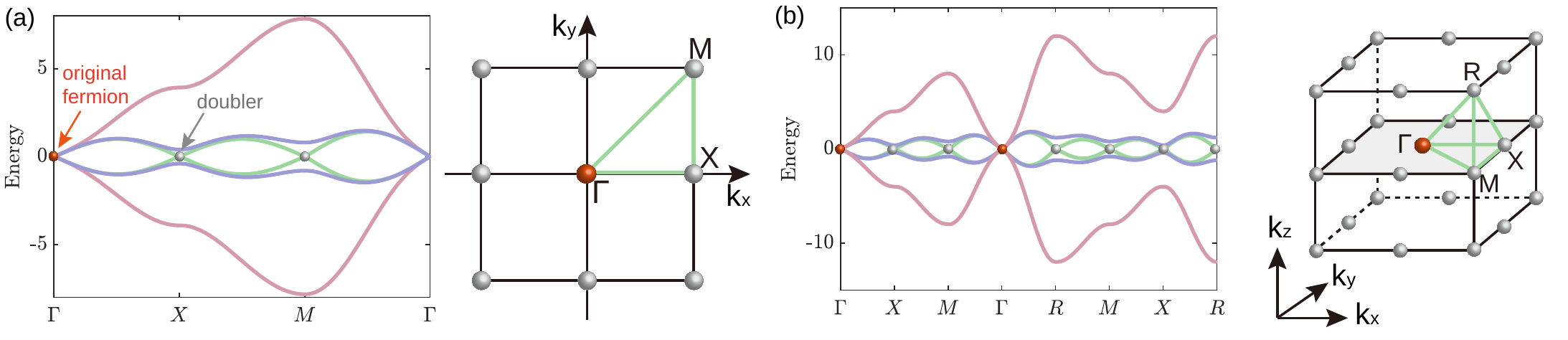}\\
  \caption{(a)(b) The energy spectra and the first Brillouin zones of 2D and 3D systems. Here, we set $a=1$ and $\hbar v=1$. The green, blue, and pink curves correspond to the parameters $b=0$, 0.1, and 1, respectively. The original fermion and doublers are labeled by red and gray dots in the first Brillouin zones, respectively.}\label{BS}
\end{figure}

\section{II. The propagation of DW bound state}
In this section, we show how to obtain the solution of the DW bound state and the propagation of the DW states. Considering the DW of Fig. 4(e) in the main text with periodic boundary condition in $\hat{y}$ direction, we can write the secular equation near the DW as \cite{Shenbook,Shan2010}
\begin{equation}\label{SE}
[2\omega C_1(-i\partial_x\sigma_x+k_y\sigma_y)+\omega C_2(x)(-\partial^2_x+k_y^2)\sigma_z]\phi(x,y)=J\phi(x,y)
\end{equation}

\noindent The solutions of Eq. \eqref{SE} are given by
\begin{equation}
\begin{aligned}
&J=2\omega C_1k_y,\\
&\phi(x,y)=\chi_y\sqrt{\lambda_2(k_y)}\exp[-\lambda_2(k_y) \vert x\vert+ik_yy],
\end{aligned}
\end{equation}
with $\chi_y=\frac{\sqrt{2}}{2}(-i,1)^{\rm T}$ and $\lambda_2(k_y)=\frac{C_1}{\vert C_2 \vert}+\sqrt{(\frac{C_1}{C_2})^2+k_y^2}$. The effective velocity of the bound state is $v_{\rm eff}=\frac{\partial J}{\partial k_y}=2\omega C_1$, indicting the DW state propagating along $\hat{y}$ direction.

To demonstrate the time evolution of the DW bound state, we perform the circuit simulation with LT{\tiny SPICE}. As shown in Fig. \ref{ASM_time}(a), we consider a sample with $20\times11$ ``spins" and input a Gaussian AC signal $i(t)=I_0\exp[-\frac{(t-t_0)^2}{(\Delta t)^2}]\sin[\omega(t-t_0)]$ close to the DW indicated by the arrow in the first subfigure of Fig. \ref{ASM_time}(a). Here, we set $I_0=1$ mA, $\Delta t=20$ $\mu$s, and $t_0=50$ $\mu$s. Then, we plot the voltage propagation at different moments and observe the signal propagation along the (DW channel) $\hat{y}$ direction. When the signal arrives at the sample edge, it will leak to bulk nodes and form a loop, as shown in the fourth subfigure of Fig. \ref{ASM_time}(a). Finally, the signal becomes a steady bound state. We also calculate the steady-state voltage by the formula $V={\mathcal J}^{-1}I$, and the result is plotted in Fig. \ref{ASM_time}(b), which is consistence with the theoretical calculations.

\begin{figure*}[h!]
  \centering
  \includegraphics[width=1\textwidth]{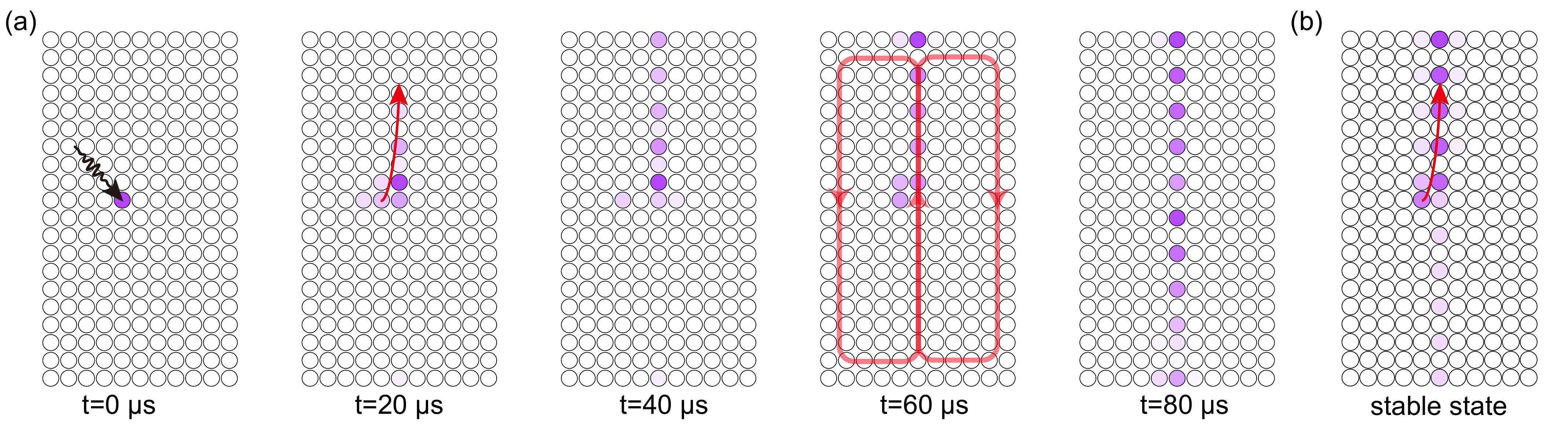}\\
  \caption{(a) The propagation of the DW state with the black arrow indicating the position of the signal source. (b) The steady-state voltage.}\label{ASM_time}
\end{figure*}

\section{III. Experimental details}
We implement the circuit experiment on a printed circuit broad shown in Fig. \ref{PCB}(a). The circuit is composed of $10\times11$ cells, with each cell containing two nodes. The details of the circuit components are shown in the inset of Fig. \ref{PCB}(a). Figure \ref{PCB}(b) displays the experimental instruments: DC power supply (IT6332A) and impedance analyzer (E4990A), which are used to provide the power for the operational amplifiers and measure the impedance over the sample, respectively.

\begin{figure}[h!]
  \centering
  \includegraphics[width=0.9\textwidth]{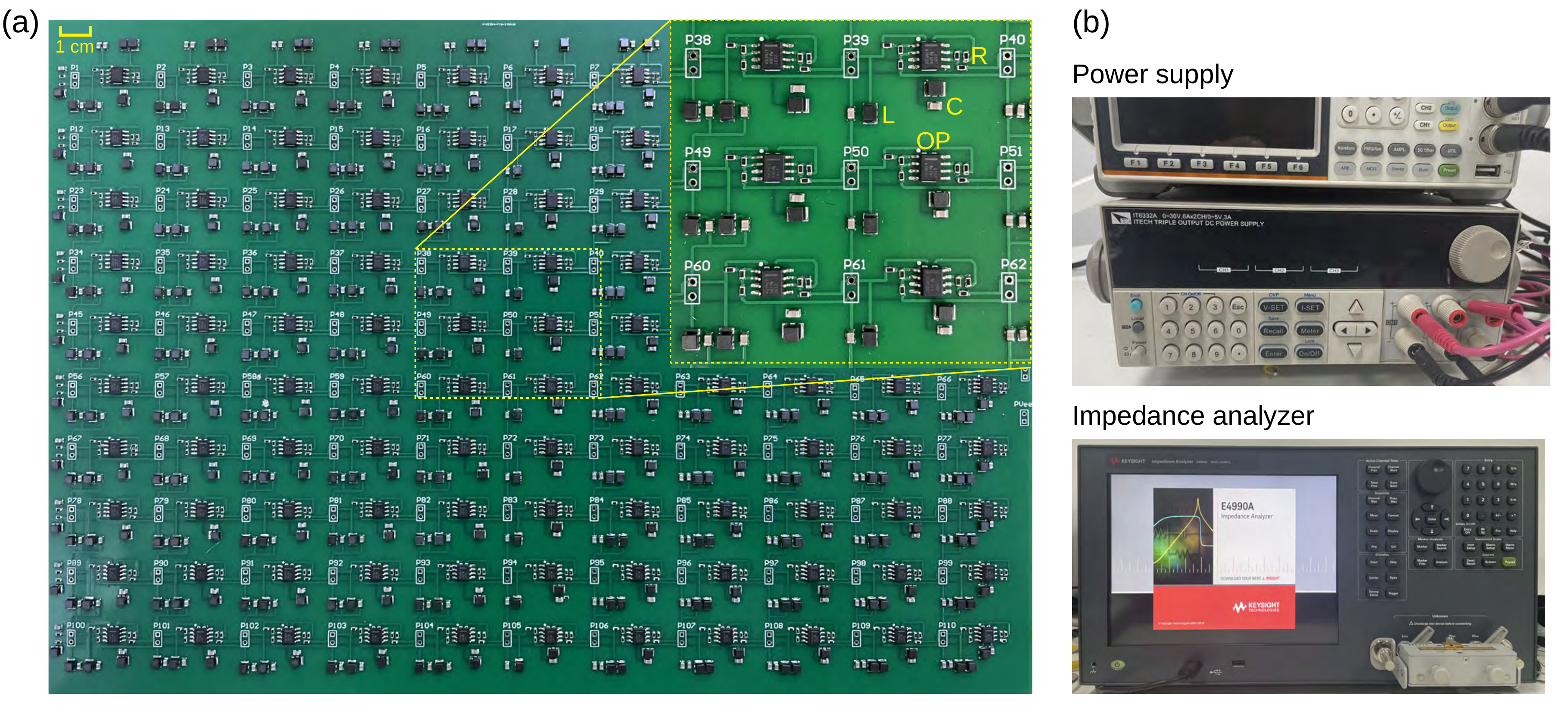}\\
  \caption{(a) The full image of the experimental printed circuit board. (b) The photos of the DC power supply and the impedance analyzer.}\label{PCB}
\end{figure}

In Table \ref{tb}, we list all elements used in our experiments, including the product companies, packages, mean values and their tolerances.

\begin{table}[h!]
\centering
\caption{Electric elements used in experiments.}
\begin{tabular}{ccccc}
\hline
Electric elements  & Company & packages  &mean value  & tolerance \\
\hline
$C_1,C_2$  & Samsung     &0805    & 1  nF         & $\pm5\%$ \\
$L$        & muRata      &1210    & 39 $\mu$H     & $\pm5\%$ \\
$R$        & Panasonic   &0603    & 200  $\Omega$ & $\pm1\%$ \\
OP         & Texas instruments          &SOIC-8  & /            & / \\
\hline
\end{tabular}
\label{tb}
\end{table}

\section{IV. Chern insulators}
To study the Chern insulator carefully, we consider two insulating phases with $\Delta=-1.5$ and $-0.5$. In Figs. \ref{CI}(a) and \ref{CI}(b), we display the admittance spectra for a ribbon with infinite size in $\hat{x}$ direction and $\mathcal{N}_y=50$ nodes in $\hat{y}$ direction. In the band gap, one can see the crossing of two spectra, manifesting as two chiral edge modes along the two boundaries. Next, we consider a finite-size square lattice with $10\times20$ ``spins" ($20\times20$ nodes) to study the chiral edge mode. The admittance spectrum is given in Fig. \ref{CI}(c) with the wave function near $j_n=0~\Omega^{-1}$ and the impedance distribution plotted in the insets. With the same simulation method, we obtain the voltage propagation of the Chern insulator at different moments and observe a chiral edge mode [see Fig. \ref{CI}(d)].

\begin{figure}[h!]
  \centering
  \includegraphics[width=0.98\textwidth]{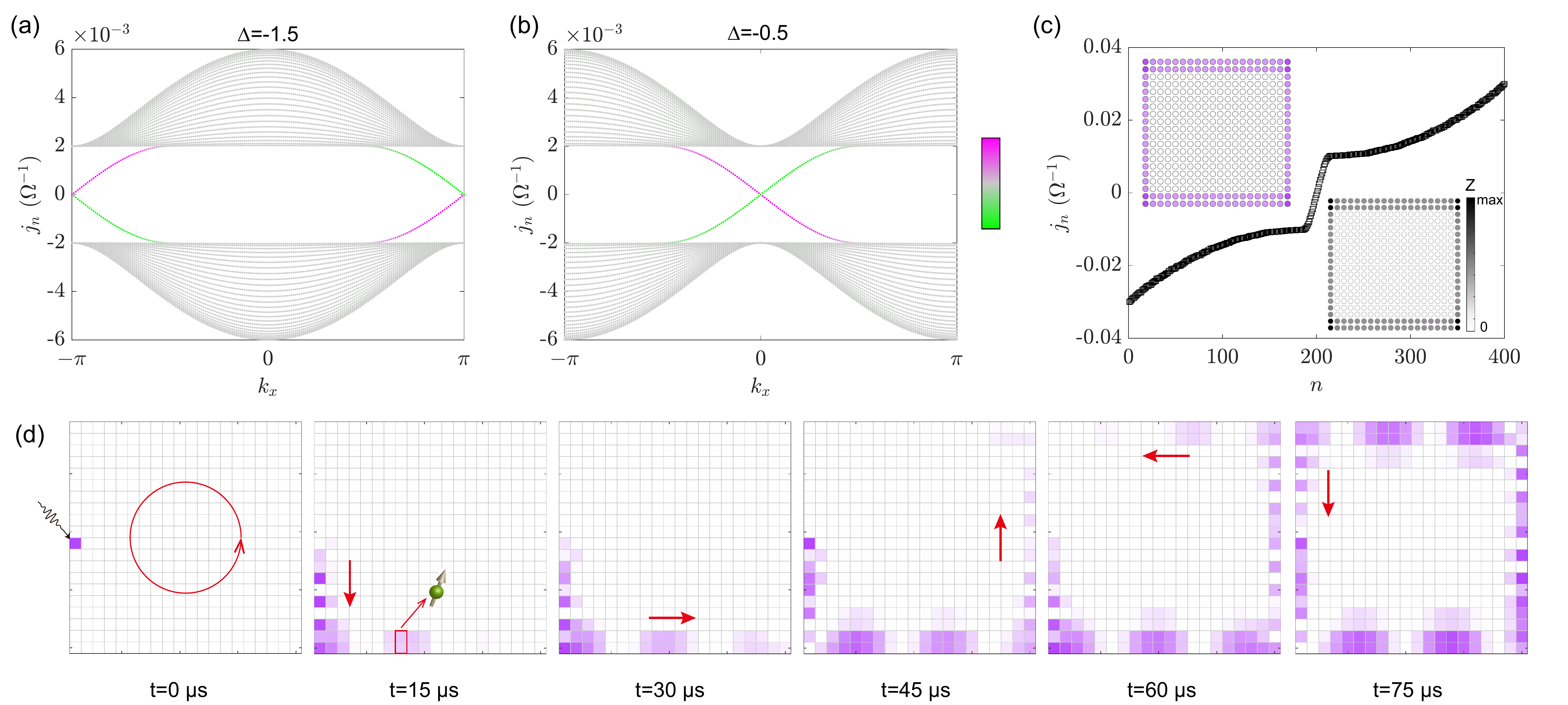}\\
  \caption{(a),(b) The band structures of the ribbon configurations. The magenta and green spectra represent the chiral boundary modes localized at the top and bottom edges, respectively. (c) The admittance spectrum with the insets showing the wave function near $j_n=0~\Omega^{-1}$ (top left corner) and the impedance distribution (bottom right corner). (d) The time evolution of the topological boundary modes. The red arrows indicate the propagation directions of the voltage signals.}\label{CI}
\end{figure}

\section{V. Three-dimensional Wilson fermions}

\begin{figure*}[t!]
  \centering
  \includegraphics[width=0.9\textwidth]{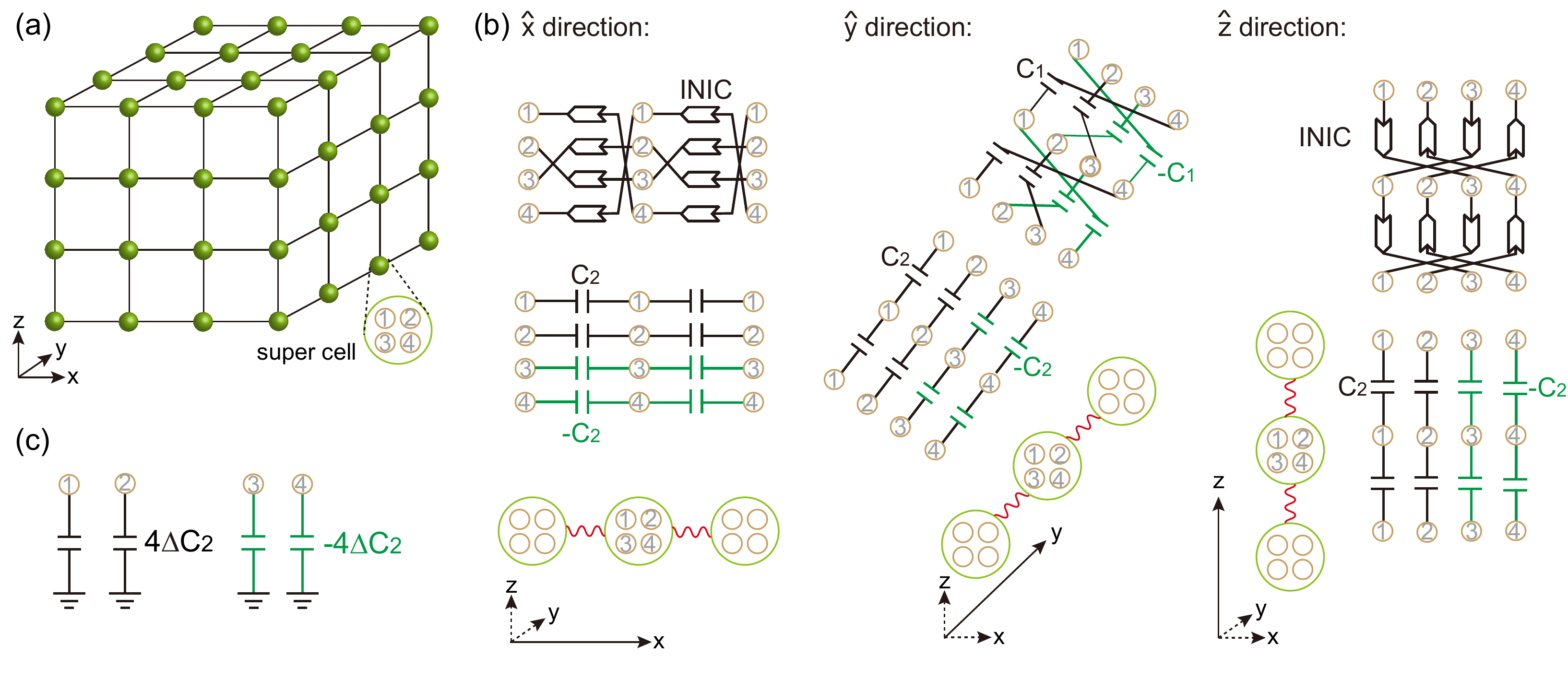}\\
  \caption{(a) Three-dimensional hyper-cubic lattice model with four sites in each supercell. (b) The interactions between two cells along $\hat{x}$, $\hat{y}$, and $\hat{z}$ directions. (c) The realization of the on-site potentials.}\label{3Dmodel}
\end{figure*}

We consider a 3D hyper-cubic lattice with four sites in each cell, as shown in Fig. \ref{3Dmodel}{\bf a}. The hopping terms and on-site potentials are shown in Fig. \ref{3Dmodel}{\bf b} and Fig. \ref{3Dmodel}{\bf c}, respectively. One can write the circuit Laplacian as
\begin{equation}
\begin{aligned}
\hat{x}:~&j_{14}=j_{23}=j_{32}=j_{41}=-G\exp(-ik_x)+G\exp(ik_x)=2iG\sin k_x,\\
&j_{11}=j_{22}=2i\omega C_2-i\omega C_2\exp(-ik_x)-i\omega C_2\exp(ik_x)
=2i\omega C_2(1-\cos k_x),\\
&j_{33}=j_{44}=-2i\omega C_2+i\omega C_2\exp(-ik_x)+i\omega C_2\exp(ik_x)=-2i\omega C_2(1-\cos k_x).
\end{aligned}
\end{equation}

\begin{equation}
\begin{aligned}
\hat{y}:~&j_{14}=j_{32}=i\omega C\exp(-ik_y)-i\omega C\exp(ik_y)=i\omega C_1(-2i\sin k_y),\\
&j_{23}=j_{41}=-i\omega C\exp(-ik_y)+i\omega C\exp(ik_y)=i\omega C_1(2i\sin k_y),\\
&j_{11}=j_{22}=2i\omega C_2-i\omega C_2\exp(-ik_y)-i\omega C_2\exp(ik_y)=2i\omega C_2(1-\cos k_y),\\
&j_{33}=j_{44}=-2i\omega C_2+i\omega C_2\exp(-ik_y)+i\omega C_2\exp(ik_y)=-2i\omega C_2(1-\cos k_y).
\end{aligned}
\end{equation}

\begin{equation}
\begin{aligned}
\hat{z}:~&j_{13}=j_{31}=-G\exp(-ik_z)+G\exp(ik_z)=2iG\sin k_z,\\
&j_{24}=j_{42}=G\exp(-ik_z)-G\exp(ik_z)=-2iG\sin k_z,\\
&j_{11}=j_{22}=2i\omega C_2-i\omega C_2\exp(-ik_z)-i\omega C_2\exp(ik_z)=2i\omega C_2(1-\cos k_z),\\
&j_{33}=j_{44}=-2i\omega C_2+i\omega C_2\exp(-ik_z)+i\omega C_2\exp(ik_z)=-2i\omega C_2(1-\cos k_z).
\end{aligned}
\end{equation}

Summarizing the above equations, we obtain
\begin{equation}
\mathcal{J}(\omega)=i\left(
              \begin{array}{cccc}
                j_0 & 0 & 2G\sin k_z & 2G\sin k_x-2i\omega C_1\sin k_y \\
                0 & j_0 & 2G\sin k_x+2i \omega C_1\sin k_y & -2G\sin k_z \\
                2G\sin k_z & 2G\sin k_x-2i\omega C_1\sin k_y & -j_0 & 0 \\
                2G\sin k_x+2i\omega C_1\sin k_y & -2G\sin k_z & 0 & -j_0 \\
              \end{array}
            \right),
\end{equation}
with $j_0=4\omega C_2(\sin^2\frac{k_x}{2}+\sin^2\frac{k_y}{2}+\sin^2\frac{k_z}{2})$.

Similarly, if expressing $\mathcal{J}(\omega)=i\mathcal{H}(\omega)$, one can obtain the tight-binding Hamiltonian
\begin{equation} \label{HD}
\mathcal{H}(\omega)=2G\sin(k_x)\alpha_x+2\omega C_1\sin(k_y)\alpha_y
+2G\sin(k_z)\alpha_z+4
\omega C_2[\sin^2\frac{k_x}{2}+
\sin^2\frac{k_y}{2}+\sin^2\frac{k_z}{2}]\beta,
\end{equation}
where $\alpha_x=\sigma_x\otimes\sigma_x=\left(
                                          \begin{array}{cccc}
                                            0 & 0 & 0 & 1 \\
                                            0 & 0 & 1 & 0 \\
                                            0 & 1 & 0 & 0 \\
                                            1 & 0 & 0 & 0 \\
                                          \end{array}
                                        \right)$,
$\alpha_y=\sigma_x\otimes\sigma_y=\left(
                                          \begin{array}{cccc}
                                            0 & 0 & 0 & -i \\
                                            0 & 0 & i & 0 \\
                                            0 & -i & 0 & 0 \\
                                            i & 0 & 0 & 0 \\
                                          \end{array}
                                        \right)$,
$\alpha_z=\sigma_x\otimes\sigma_z=\left(
                                          \begin{array}{cccc}
                                            0 & 0 & 1 & 0 \\
                                            0 & 0 & 0 & -1 \\
                                            1 & 0 & 0 & 0 \\
                                            0 & -1 & 0 & 0 \\
                                          \end{array}
                                        \right)$,
and
$\beta=\sigma_z\otimes\sigma_0=\left(
                                          \begin{array}{cccc}
                                            1 & 0 & 0 & 0 \\
                                            0 & 1 & 0 & 0 \\
                                            0 & 0 & -1 & 0 \\
                                            0 & 0 & 0 & -1 \\
                                          \end{array}
                                        \right)$.

The energy spectra are given by
\begin{equation}
j_n=\pm\sqrt{j_0^2+j_x^2+j_y^2+j_z^2},
\end{equation}
with $j_0=4\omega C_2[\sin^2\frac{k_x}{2}+
\sin^2\frac{k_y}{2}+\sin^2\frac{k_z}{2}]$, $j_x=2G\sin k_x$, $j_y=2\omega C_1\sin k_y$, and $j_z=2G\sin k_z$.

%
%

\begin{figure}
  \centering
  \includegraphics[width=0.8\textwidth]{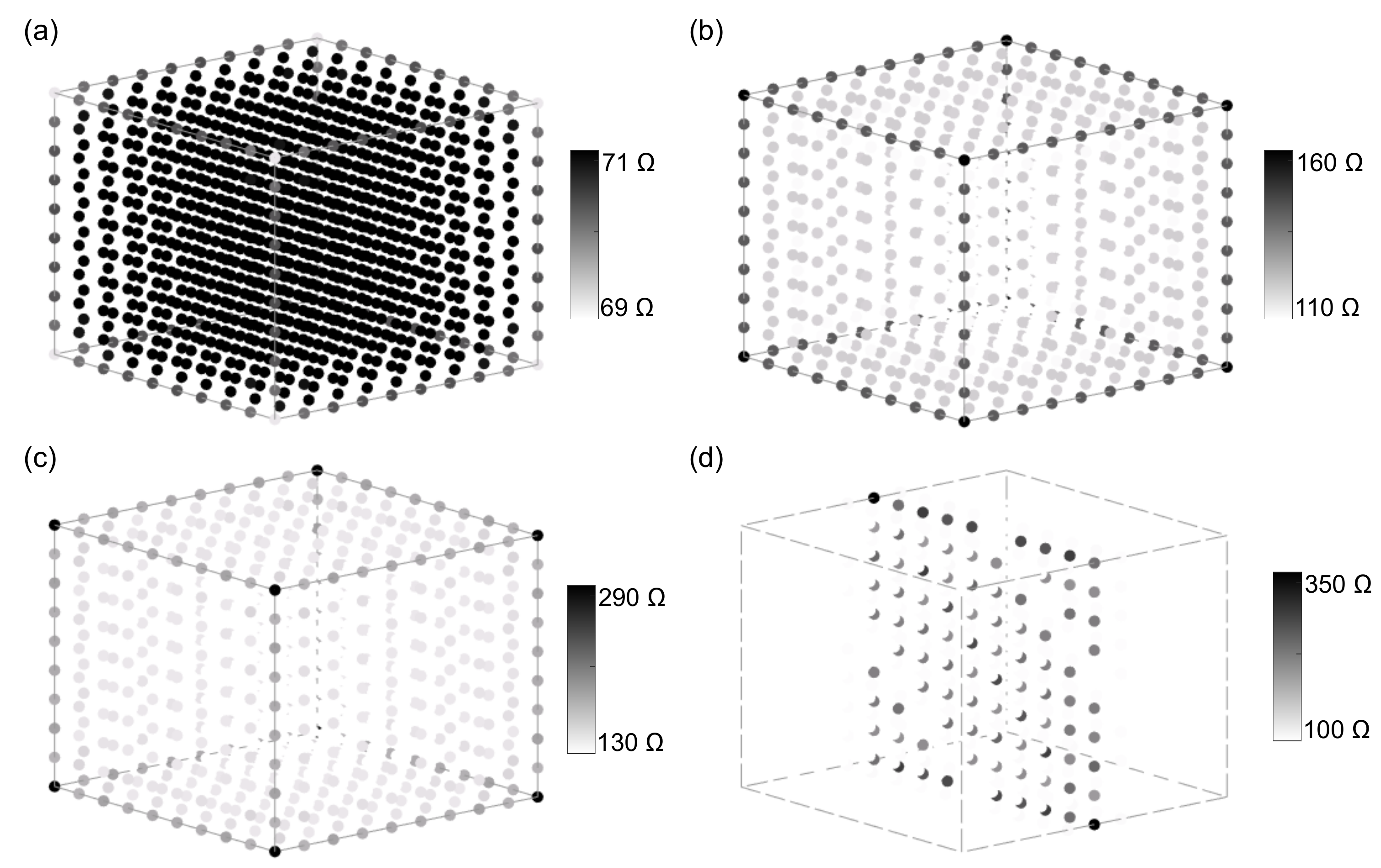}\\
  \caption{(a)-(d) The distributions of impedance for the quantum anomalous semimetal state, topological insulator state, semimetal state, and domain wall state. }\label{3DAS}
\end{figure}

One can rewrite the Eq. \eqref{HD} as
\begin{equation}
\mathcal{H}(\omega)=\left(
              \begin{array}{cc}
                j_0 & {\bf j}\cdot{\bm \sigma} \\
                {\bf j}\cdot{\bm \sigma} & -j_0 \\
              \end{array}
            \right),
\end{equation}
with ${\bf j}=(j_x,j_y,j_z)$.

Due to the presence of global sublattice symmetry $\Gamma \mathcal{H}\Gamma^{-1}=-\mathcal{H}$ with $\Gamma=\exp[-i\frac{\pi}{4}(d+1)]\beta\prod_{i=1}^d\alpha_i$, the above Hamiltonian can be expressed as the block off-diagonal form
\begin{equation}
\mathcal{H'}(\omega)=\left(
              \begin{array}{cc}
                0 &  q(\bf{k})\\
                q(\bf{k})^\dagger & 0 \\
              \end{array}
            \right),
\end{equation}
with $q(\bf{k})={\bf j}\cdot{\bm \sigma}$$-ij_0$.
To characterize the topological properties, we evaluate the 3D winding number \cite{Schnyder2008} as
\begin{equation}
w_3=-\frac{1}{24\pi^2}\int_{\rm BZ}{\rm trace}[(q^{-1}\partial_{k_x} q)(q^{-1}\partial_{k_y} q)(q^{-1}\partial_{k_z} q)].
\end{equation}

Next, we consider a finite-size sample with $10\times10\times10$ ``spin" ($4000$ nodes). As shown in Figs. \ref{3DAS}(a)-(c), we show the impedance of the sample for the quantum anomalous semimetal phase, topological insulator phase, semimetal phase, respectively. These results resemble the 2D cases. To show the bulk-boundary correspondence, we also form a 2D domain wall along $\hat{x}$ direction (with $11\times10\times10$ ``spin") and observe the surface state confined inside the DW, as shown in Fig. \ref{3DAS}(d).

\end{widetext}

\end{document}